\documentclass{osa-article}

\journal{oe}
\usepackage{soul}
\makeatletter
\newcommand{\rmnum}[1]{\romannumeral #1}
\newcommand{\Rmnum}[1]{\expandafter\@slowromancap\romannumeral #1@}
\makeatother

\begin{document}
\title{Noise analysis of the atomic superheterodyne receiver based on flat-top laser beams}

\author{Zheng Wang,\authormark{1,2,$\ddagger$} Mingyong Jing,\authormark{1,2,3,$\ddagger$,*} Peng Zhang,\authormark{1,2} Shaoxin Yuan,\authormark{1,2} Hao Zhang,\authormark{1,2} Linjie Zhang,\authormark{1,2} Liantuan Xiao,\authormark{1,2} and Suotang Jia\authormark{1,2}}

\address{\authormark{1}State Key Laboratory of Quantum Optics and Quantum Optics Devices, Institute of Laser Spectroscopy, Shanxi University, Taiyuan, Shanxi 030006, China\\
\authormark{2}Collaborative Innovation Center of Extreme Optics, Shanxi University, Taiyuan, Shanxi 030006, China\\
\authormark{3}State Key Laboratory of Precision Measurement Technology and Instruments, Department of Precision Instrument, Tsinghua University, Haidian, Beijing 100084, China\\
\authormark{$\ddagger$}These authors contributed equally to this work.}

\email{\authormark{*}Corresponding author: jmy@sxu.edu.cn} 



\begin{abstract}
Since its theoretical sensitivity is limited by quantum noise, radio wave sensing based on Rydberg atoms has the potential to replace its traditional counterparts with higher sensitivity and has developed rapidly in recent years. However, as the most sensitive atomic radio wave sensor, the atomic superheterodyne receiver lacks a detailed noise analysis to pave its way to achieve theoretical sensitivity. In this work, we quantitatively study the noise power spectrum of the atomic receiver versus the number of atoms, where the number of atoms is precisely controlled by changing the diameters of flat-top excitation laser beams. The results show that under the experimental conditions that the diameters of excitation beams are less than or equal to 2 mm and the read-out frequency is larger than 70 kHz, the sensitivity of the atomic receiver is limited only by the quantum noise and, in the other conditions, limited by classical noises. However, the experimental quantum-projection-noise-limited sensitivity this atomic receiver reaches is far from the theoretical sensitivity. This is because all atoms involved in light-atom interaction will contribute to noise, but only a fraction of them participating in the radio wave transition can provide valuable signals. At the same time, the calculation of the theoretical sensitivity considers both the noise and signal are contributed by the same amount of atoms. This work is essential in making the sensitivity of the atomic receiver reach its ultimate limit and is significant in quantum precision measurement.
\end{abstract}

\section{Introduction}
Radio wave measurement based on Rydberg atoms has attracted widespread attention recently due to advantages such as SI traceable, ultra-broadband response, and potentially ultra-high sensitivity, and has extensive applications in antenna calibration\cite{antenna}, terahertz wave sensing \cite{terachertz}, sub-wavelength imaging\cite{subwavelength}, and ultra-broadband communication\cite{ultra}. The atomic radio wave sensor's most attractive advantage is that only the quantum projection noise limits its theoretical sensitivity and can beat its classical counterpart\cite{sensitivitiy}. However, as the most advanced atomic radio wave sensor, the atomic superheterodyne receiver's (atomic superhet)\cite{jing2020} sensitivity is still far from the theoretical expectation\cite{1-400}. Therefore, detailed noise analysis of atomic superhet is needed to pave its way to achieve theoretical sensitivity.

Thanks to the rapid development of atomic clocks and magnetometers, many works have been done on the noise analysis of light-atom interaction systems. In these works, the noise power spectrum (NPS) has become a powerful noise analysis tool. Based on the analysis of NPS, the typical noise sources in light-atom interaction systems are as follows. The first type of noise comes from the conversion of laser phase or frequency noise to amplitude noise caused by the nonlinear coherence effect of the atomic medium in response to the fluctuating field\cite{6,07,12,22}. This noise has been widely studied in absorption\cite{2,10,11,14,21,26} and electromagnetically induced transparency (EIT)\cite{8,15,20,25} spectroscopies. The second type of noise comes from the random perturbation of laser light scattering through the atomic medium due to dephasing caused by spontaneous emission or transit of atoms\cite{4,13,16,18,19}. The third type of noise is quantum projection noise, arising from the random projection of the state vector into one of the states compatible with the measurement process and is an inherent feature of quantum mechanics\cite{24,27,28}. Unfortunately, all these kinds of noises mixed in the output spectrum of atomic superhet, making it very difficult to trace the source. Therefore, in addition to the NPS, extra methods are needed to provide a clear trace of the noise.
 
In order to trace the noise source of the atomic superhet, this work first realizes a precise control of the number of atoms involved in the measurement by adjusting the size of the flat-top excitation laser beams. It then comprehensively characterizes the output noise of the atomic superhet based on the NPS. The noise at different read-out frequencies is analyzed separately, and the power-law scaling between noise power and the number of atoms is finally explored. This work clarifies the noise sources in the current atomic superhet and lights the way for further sensitivity improvements.

\section{Experimental setup}

\begin{figure}[ht!]
	\includegraphics[width=5in]{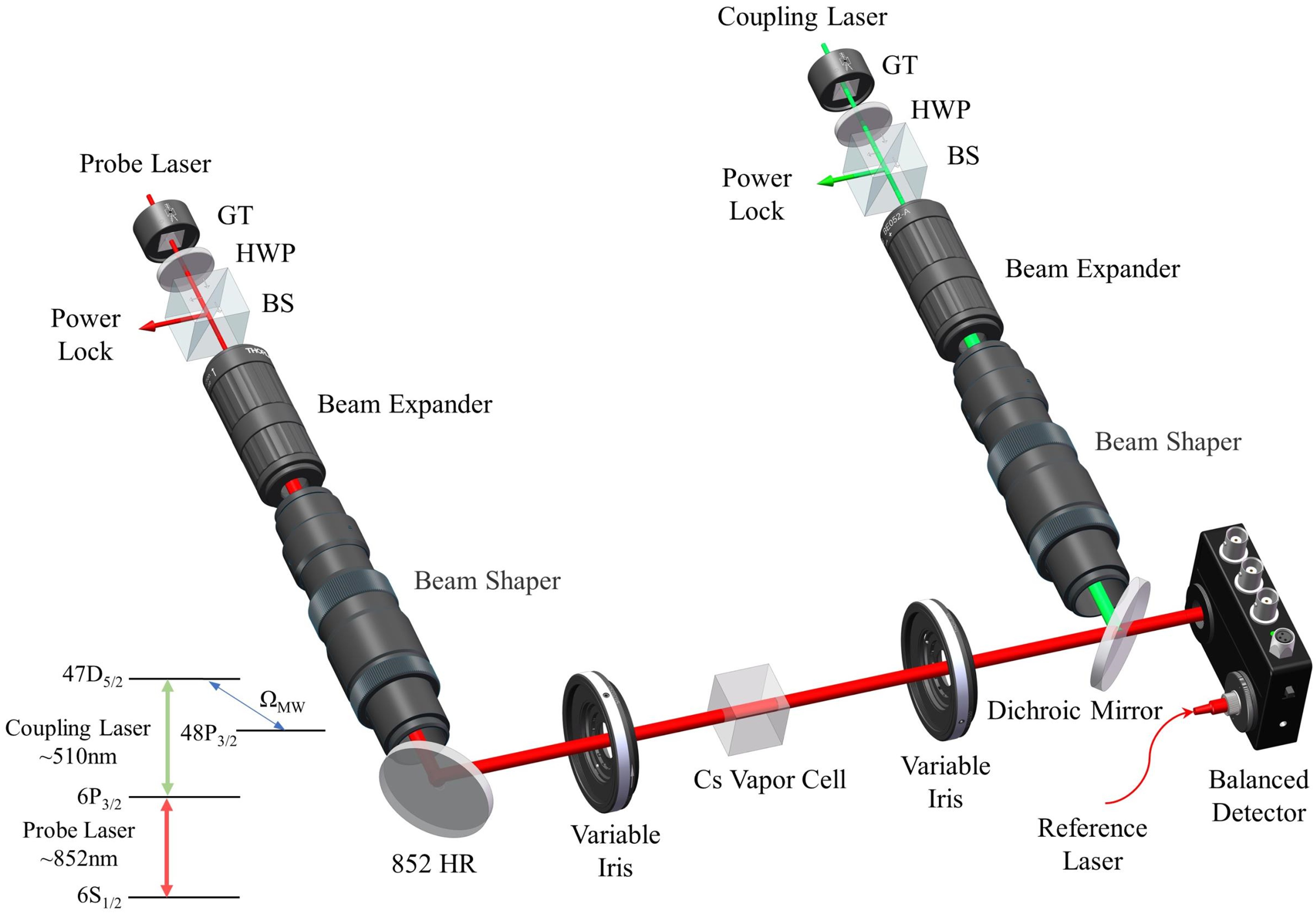}
 \centering
\caption{\label{figure1}Experimental setup. GT: Glan-Taylor Polarizer, HWP: Half-Wave Plate, BS: Beam Splitter (T:R=9:1), and HR: High Reflection dielectric mirror. Inset: Energy level diagram.}
\end{figure}

Figure. \ref{figure1} shows the experimental setup and energy level diagram (inset). We use the same energy level structure as our previous work on atomic superhet\cite{jing2020}, except that the $47\rm{D} _{5/2}\rightarrow 48\rm{P} _{3/2}$ transition is only coupled by the local microwave (MW) field since this work focuses on the noise of the atomic superhet. The local MW field is incident into the atomic vapor cell with vertical polarization and propagates perpendicular to the propagation direction of the probe and coupling laser beams, with a Rabi frequency of about $2\pi \times 10.68$ MHz. Multipath propagation effect of microwave has been suppressed by placing microwave-absorbing materials around the atomic receiver and experiment platform. The probe and coupling laser systems used in this work are the same as our previous work, ensuring a low noise feature of their frequencies. The probe and coupling lasers are coupled from polarization-maintaining fibers into free space through collimators. High purity vertical polarization of excitation beams is then achieved using Glan Taylor polarizers with more than 60 dB polarization extinction ratio and half-wave plates, which ensures that only $\pi$ transitions can be excited. After polarization purification, probe and coupling lasers are split out 10\% of their through the unpolarized beam splitters for power monitoring and locking, thereby reducing their intensity noise. The lasers with low frequency and intensity noise help us achieve precision measurements of NPS for atomic superhet.

To trace the noise's source by verifying the power-law scaling of its power versus atomic number, the total number of atoms should be adjustable, which, in this work, is achieved by varying the diameters of flat-top beams to change the interaction volume while keeping the atomic number density constant. The realization of size-variable flat-top laser beams is described as follows. The power-stabilized lasers are first expanded to a suitable beam size (6.1-6.2 mm for 510 beam and 6.3-6.4 mm for 852 beam) by adjustable beam expansions and then converted from Gaussian profile to flat-top profile by $\pi$-shaper beam shapers. By optimizing the sizes and positions of the laser beams entering the beam shapers and parameters of beam shapers, we pre-compensated the wavefront differences of laser beams caused by their reflection through the dielectric mirror (for probe laser) or the dichroic mirror (for coupling laser), ensuring they have proper flat-top distribution when entering the vapor cell. The uniform intensity distributions of flat-top beams ensure that the average Rabi frequencies ($2\pi \times 5.13$ MHz for probe laser and $2\pi \times 0.29$ MHz for coupling laser) of light-atom interaction maintain constant when adjusting the beam diameters, achieve a linear relationship between beam size and the total number of atoms. Finally, the sizes of flat-top beams are adjusted by variable irises. Each variable irises consists of a set of custom-made fixed-size pinholes of varying sizes and a motorized rotation stage. The motorized rotation stages are driven by resonant piezoelectric motors technology and have a repeatability of up to 1.75 mrad, which ensures high reproducibility of the experiment and reduces the measurement error while exchanging different size pinholes. Based on these devices, flat-top beams with a diameter of up to 6.4 mm can be achieved.

Eventually the flat-top probe and coupling lasers propagate in the opposite direction and coincide in the cesium vapor cell with inner dimensions of $25\times25\times25$ mm$^3$. Then the probe laser transmission is detected by one of the balanced detector's optical input ports (i.e., signal port). A reference 852 nm laser is input to the other optical port (i.e., reference port) of the balanced detector to cancel the DC offset of photocurrent, which can reduce the photodetector amplifier's saturation and extend the dynamic range of the balanced detector. A spectrum analyzer analyzes the balanced electrical signal outcome by the balanced detector to obtain the NPS.

\section{Experimental results and discussions}

\begin{figure}[ht!]
\includegraphics[width=4in]{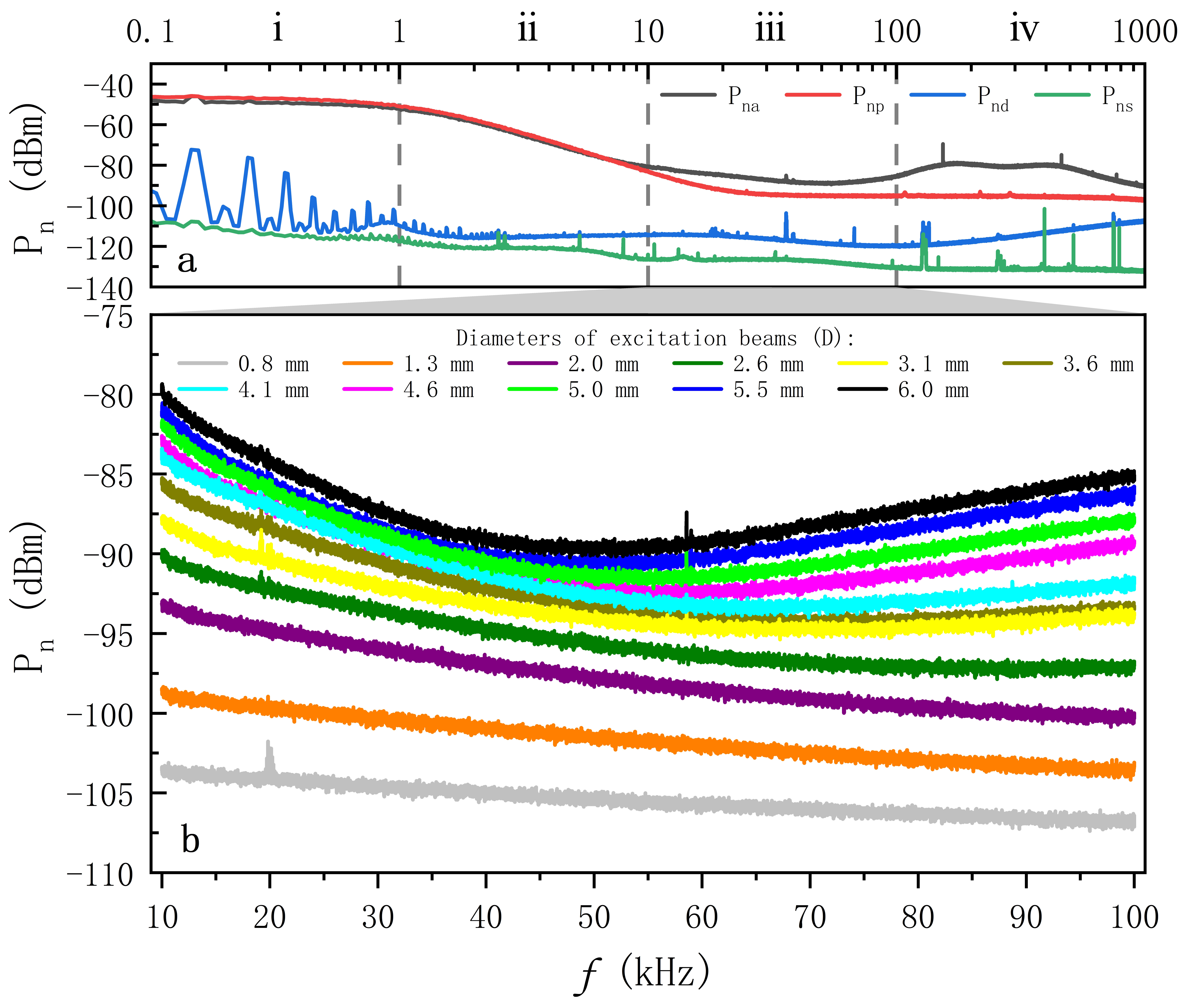}
\centering
\caption{\label{figure2} The NPSs of atomic superhet for different flat-top beam diameters. The intensity of the probe beam is kept at $31.85~\mu\rm{W}/\rm{mm} ^{2}$ in this work, corresponding to a Rabi frequency of $2\pi\times5.13$ MHz, while the intensity of the coupling beam is kept at $1.92~\rm{mW}/\rm{mm}^{2}$, corresponding to a Rabi frequency of $2\pi\times0.29$ MHz. The frequency of the local MW field is 6.95 GHz, with an electric field strength of $0.41~ \rm{V/m}$ (RMS value), corresponding to a Rabi frequency of $2\pi\times10.68$ MHz. a). The NPS of atomic superhet from 0.1 kHz to 1000 kHz. X-axis ($f$) means read-out frequency, and Y-axis ($\rm{P_n}$) means noise power. The black curve ($\rm{P_{na}}$) is the NPS of the atomic superhet with 6 mm diameter excitation beams. The red curve ($\rm{P_{np}}$) is the NPS of the probe laser, which is measured by removing the atomic vapor cell and keeping the probe beam's diameter and power input to the signal port of the balanced detector as the same as $\rm{P_{na}}$. The blue curve ($\rm{P_{nd}}$) is the NPS of the balanced detector without any optical input. The green curve ($\rm{P_{ns}}$) is the NPS of the spectrum analyzer. b). The NPSs of atomic superhet for different flat-top beam diameters from 10 kHz to 100 kHz.}
\end{figure}

We first measured the NPSs of atomic superhet with different excitation beam diameters, shown in Fig. \ref{figure2}. Figure. \ref{figure2}(a) compares noises between the atomic superhet with 6 mm diameter excitation beams ($\rm{P_{na}}$), probe laser ($\rm{P_{np}}$), and measurement devices ($\rm{P_{nd}}$ for balanced detector and $\rm{{P_{ns}}}$ for spectrum analyzer). The noises of measurement devices contribute less to the read-out noise of atomic superhet as they are at least 10 dB lower than it. The noise of the atomic superhet is divided into four regions according to its property. In regions \rmnum{1}~(0.1-1 kHz) and \rmnum{2}~(1-10 kHz), the atomic superhet's noise mainly comes from the noise of the probe laser. It shows a white noise characteristic in region \rmnum{1}, whoes power keeps constant with read-out frequency, and a $1/f^n~(1/f^3~\&~1/f^4)$\cite{freview} noise characteristic in region \rmnum{2}, whoes power is inversely proportional to the nth power of read-out frequency. The noises in regions \rmnum{1}~and \rmnum{2}~mainly come from the laser system's thermal noise and the measurement system's mechanical jitter. The noise in regions \rmnum{3}~and \rmnum{4}~mainly comes from light-atom interaction. In region \rmnum{4}~(100 kHz- 1 MHz), the frequency noise generated by servo resonances occurring near the servo bandwidth of the laser frequency servo systems is converted to read-out noise of atomic superhet through the nonlinear coherence effect of the atomic medium. The noise in region \rmnum{3}~(10 kHz-100 kHz) is more complicated than others; thus,  a series of noise powers with different interacting atom numbers are measured to trace the noise source, shown in Fig. \ref{figure2}(b). The noise power of atomic superhet in region \rmnum{3}~increases as the atom number increases, with a different increase rate at different read-out frequencies, e.g., increasing rapidly at low and high frequencies and slow at intermediate frequencies, which means that at different read-out frequencies, it is dominated by different kinds of noise.

\begin{figure}[ht!]
\includegraphics[width=4in]{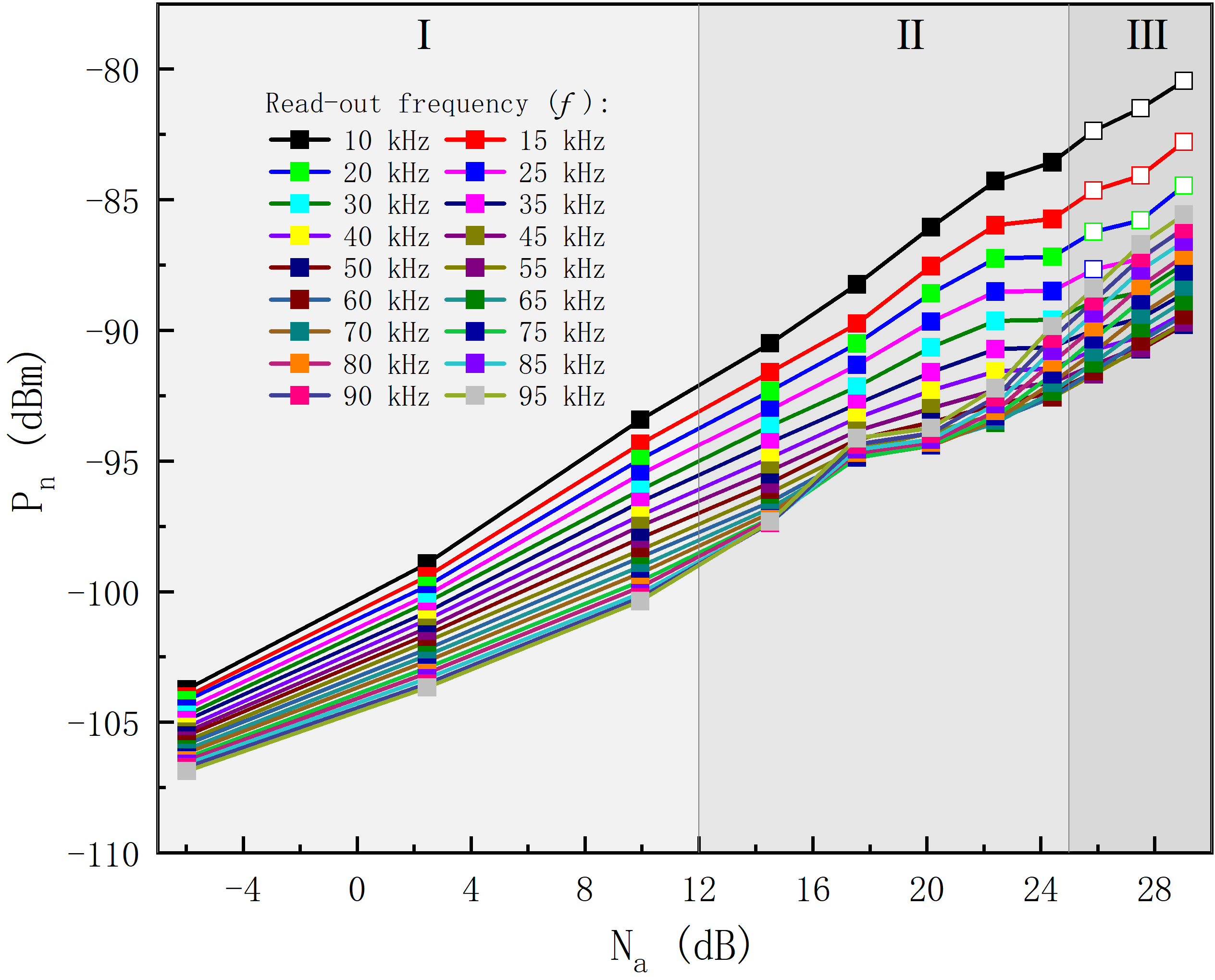}
\centering
\caption{\label{figure3} The noise power of atomic superhet ($\rm{P_n}$) versus $\rm{N_a}$ at different read-out frequencies, where $\rm{N_a}=20\times\log{\rm{A_{e}}}$ is the relative atom number relative to the number of atoms per unit area ($\rm{mm}^{2}$), and $\rm{A_{e}}=\pi(\rm{D}/2)^2$ is the cross-sectional area of the excitation beams. The hollow points in region \Rmnum{3} indicate invalid data, of which read-out frequencies are less than or comparable with the transit rates of atoms at the corresponding beam diameters (see main text).}
\end{figure}

The data in Fig. \ref{figure2}(b) is further processed to quantitative analyze the relationship between the noise power and atom number at different read-out frequencies. First, the probe laser noise is reduced in the data to ensure the remaining noise mainly comes from the light-atom interaction. Then, the data is sectioned at 1 kHz intervals and averaged within each section to reduce the measurement uncertainty. Fig. \ref{figure3} shows some of the processed data.
Since the number of interacting atoms is adjusted by changing the size of excitation beams, which also changes the time of atoms crossing the laser beams simultaneously, thus the transit relaxation rate is also changed. The transition relaxation rate $\gamma$ is related to the excitation beam diameter D and the velocity of atom $v$. For a flat-top beam with uniform distribution, the relationship is given by\cite{equ1}
\begin{equation}
   \gamma =1.13\frac{v}{\rm{D}}.
\label{equ1}
\end{equation}
Variation of the transit relaxation rate changes the spectrum distribution of the noise caused by the random transit process, which adds an undesired influencing factor to noise other than the change of atom number. This factor can be eliminated by classifying the data points into in-band, close-to-band, and out-of-band sets of data points, and only the in-band and out-of-band data sets that are less affected by the change of transit relaxation rate are valid and analyzed subsequently. The classification rule of the data points is based on a comparison of read-out frequency  ($f$) and transition relaxation rate ($\gamma$), e.g., the data points are classified as in-band if their $f$ are significantly smaller than $\gamma$, classified as close-to-band if their $f$ are comparable to $\gamma$, and classified as out-of-band if their $f$ are significantly larger than $\gamma$. For each data point, $\gamma$ is calculated according to the experimental conditions using Eq. \ref{equ1}, where the average velocity of atoms $v$ is calculated by the ARC package\cite{arc}, and is about 217 $\rm{m/s}$ at a room temperature of 22.5 $^{\circ}\rm{C}$ in this work. 

The classification results of data points according to the rule are shown in Fig. \ref{figure3}. All data points in region \Rmnum{1} are in-band data points, and all data points except for the hollow ones in region \Rmnum{3} are out-of-band data points. Data points in region \Rmnum{2} and hollow data points in region \Rmnum{3} are dropped in subsequent processing. The valid data points in regions \Rmnum{1} or \Rmnum{3} are then fitted by using the equation
\begin{equation}
  \rm{P_{n}}=\rm{A}\times\rm{N_{a}}^{2\kappa}+\rm{P_{0}},
\label{equ2}
\end{equation}
where A is a proportional coefficient, $\rm{P_0}$ is a constant term unrelated to the number of atoms, and $\kappa$ is the power-law coefficient of noise amplitude (and $2\kappa$ for noise power) with the atom numbers. $\rm{P_0}$ comes from the residual noise of the measurement system that has not been eliminated from the previous data processing process. The fitting results of $\rm{P_0}$ is about -116.5 dBm, at least 10 dB lower than the noise power of atomic superhet, and is a reasonable correction allowed by the experimental error. $\kappa$ determines whether the noise source is classical or quantum. $\kappa=1$ means that the noise power is proportional to the square of the atom number, and the noise comes from the classical noise. $\kappa=0.5$ means that the noise power is proportional to the atom number, and the noise comes from quantum projection noise. $\kappa$ between 0.5 and 1 means that both classical and quantum noise exist.
\begin{figure}[ht!]
\includegraphics[width=4in]{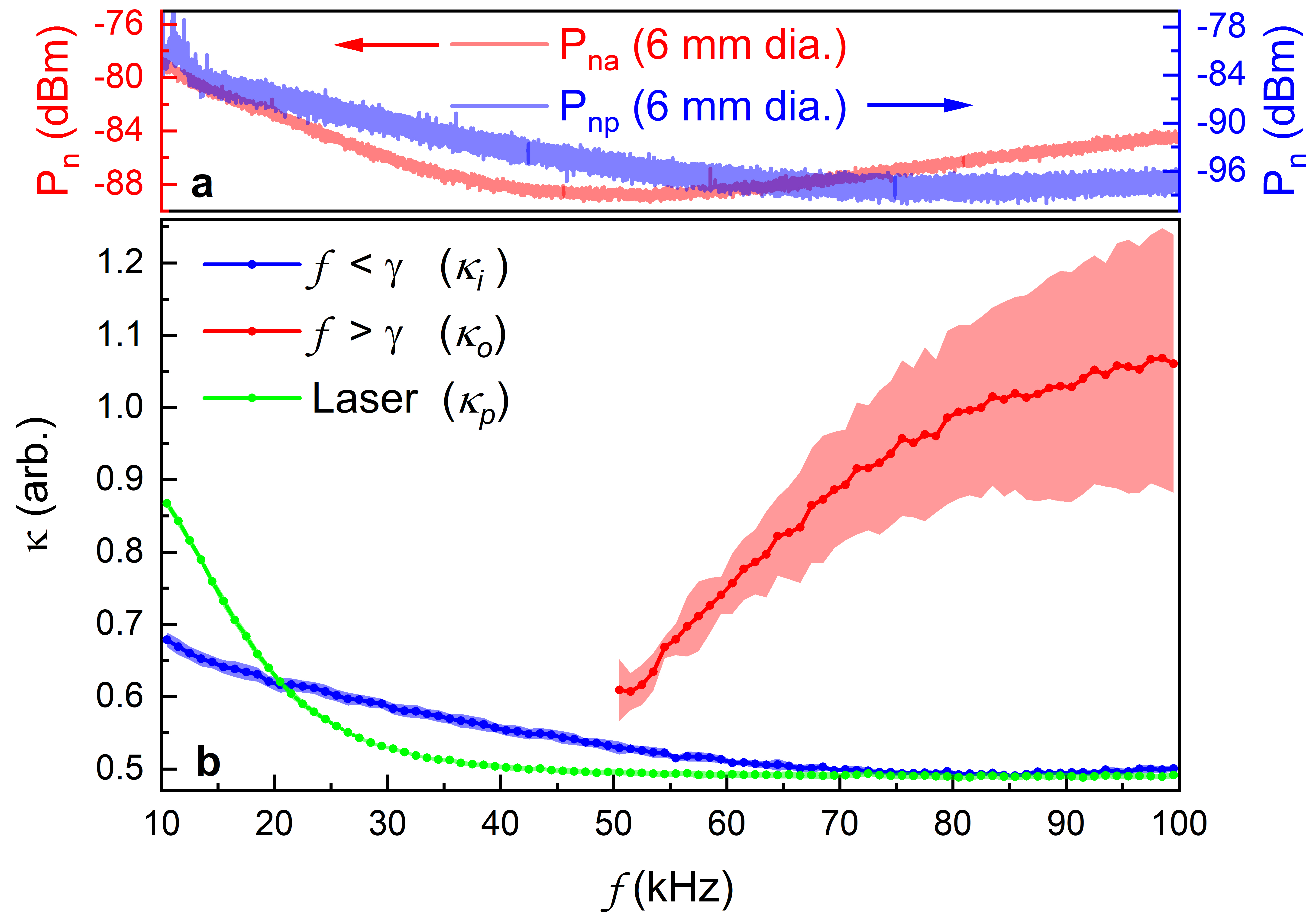}
\centering
\caption{\label{figure4} Power-law coefficient at different read-out frequencies. a). The comparison of NPS between atomic superhet and probe laser in the frequency range from 10 kHz to 100 kHz with excitation beam diameters of 6 mm. b). $\kappa$ versus $f$, the blue curve ($\kappa_i$) is the results of in-band data points. The red curve ($\kappa_o$) is the result of out-of-band data points. The green curve ($\kappa_p$) is the result of the probe laser (see main text). The error bar of data points is the statistical results of five independent experiments.}
\end{figure}

Figure. \ref{figure4}(b) shows $\kappa$ as a function of read-out frequency $f$ for in-band (blue curve, marked as $\kappa_i$) and out-of-band (red curve, marked as $\kappa_o$) data points. The $\kappa_i$ decreases as $f$ increases and converges to 0.5 when $f$ is above 70 kHz, which means under the corresponding experimental parameter ($\rm{D}\leq 2$ mm), atomic superhet achieves projection noise limit when $f$ is higher than 70 kHz. On the other hand, the $\kappa_o$ increases as $f$ increases and finally converges to 1. The reason for the opposite behaviors of $\kappa_i$ and $\kappa_o$ is that different kinds of classical noise dominate atomic superhet under corresponding parameters. For $\kappa_i$, the atomic superhet operates with small excitation beams, and the classical noise is dominated by transition noise in the frequency range of interest. The transit noise decreases gradually as $f$ increases and is significantly lower than the quantum projection noise at high frequencies, making the atomic superhet reach the quantum projection noise limit. 
For $\kappa_o$,  atomic superhet operates with excitation beams diameters larger than 5 mm, corresponding to a transit rate lower than 50 kHz. Therefore, the transit noise is mainly concentrated in data sets of which $f$ is below 50 kHz. For these data sets, the low transit rate also results in fewer valid data points than the minimum fitting required using Eq. \ref{equ2} since at least three are required for Eq. \ref{equ2} has three free parameters. 
For data sets of which $f$ is above 50 kHz, the noise of atomic superhet is dominated by a classical noise converted from laser frequency noises. The frequency noises originate from the resonances of our laser system's frequency servo loops and peak at 150 and 300 kHz for the probe and coupling laser, respectively. Thus at 50-100 kHz, this classical noise gradually rises as $f$ increases. However, this noise does not appear in the in-band set of data because it decreases faster than the quantum noise as the number of atoms decreases, and it is significantly smaller than the quantum projection noise when atomic superhet operates with small excitation beams, which corresponding a small number of interaction atoms.

In Fig. \ref{figure4}(b), We also show the power-law coefficient of the probe laser's noise with its optical power (green curve, marked as $\kappa_p$). The following procedure obtains $\kappa_p$. First, the NPSs of the probe laser at different optical powers are measured by directly inputting the probe and the reference lasers into the balance detector and measuring its balanced output with a spectrum analyzer. The probe and the reference laser have the same power $\rm{P_{ip}}$, and the change of $\rm{P_{ip}}$ is achieved by varying the probe beam diameters. The data of the probe laser's NPSs is first subtracted by the balanced detector's noise (i.e., $\rm{P_{nd}}$ in Fig. \ref{figure2}(a)), then is sectioned at 1 kHz intervals and averaged within each section. All the sectioned and averaged data points at the same read-out frequency with different optical power are fitted by equation
\begin{equation}
\rm{P_{n}}=\rm{A}\times\rm{P_{ip}}^{2\kappa_p}+\rm{P_{0}}.
\label{equ3}
\end{equation}
When $\kappa_p = 0.5$ means that the noise of the probe laser is at the shot noise limit, and $\kappa_p >0.5$ means additional classical noise exists. As a result, $\kappa_p$ shows our balanced measurement of low noise probe laser achieves shot noise limit for $f>40$ kHz. Figure. \ref{figure4}(a) also shows the NPS of atomic superhet and probe laser in a range of $10<f<100$ kHz as a reference.

\section{Conclusions }
In conclusion, we trace the noise source in atomic superhet by analyzing the power-law scaling of the noise power with atom numbers. This is achieved by measuring the NPS of atomic superhet at different interaction atom numbers, processing the data in NPS with noise reduction, sectioning and averaging, and then rearranging the processed NPS data with the same read-out frequency as a function of atom numbers to fit the power-law coefficient. Results show that in the frequency range of interested (10<$f$<100 kHz), the noise of atomic superhet is dominated by quantum projection noise at small excitation beams ($\rm{D}\leq 2$ mm) and high read-out frequency ($f>$ 70 kHz). In other cases, its noise is dominated by two kinds of classical noise: transit noise or intensity noise converted from laser frequency noise. It should be noted that although the noise of atomic superhet reaches quantum projection noise under certain experimental conditions, the sensitivity of the current atomic receiver is far from the ultimate limit predicted by theory. This is because all atoms within interaction volume provide noise, but only a small fraction (about 1/400)\cite{1-400} of them can be excited to the Rydberg state to generate signals. At the same time, the calculation of the theoretical sensitivity considers that an equivalent amount of atoms contribute to both the noise and signal. This work clarifies the noise source of the current atomic superhet and paves the way for the atomic receiver to achieve its theoretical projection-noise-limited sensitivity, e.g., the excitation ratio of Rydberg atoms in an alkali vapor cell should be increased to reach the theoretical projection-noise-limited sensitivity.
\section*{Funding}
This research is funded by the National Key R\&D Program of China (grant no. 2017YFA0304203), the National Natural Science Foundation of China (grants 12104279, 61827824 and 61975104), Shanxi Provincial Key R\&D Program (202102150101001), Science and Technology on Electronic Information Control Laboratory Fund, China-Belarus Electromagnetic Environment Effect "Belt and Road" Joint Laboratory Fund (ZBKF2022030201).

\section*{Disclosures}
The authors declare no conflicts of interest.

\bibliography{OE-manuscript}

\end{document}